\documentclass{IEEEtran}
\pdfoutput=1
\usepackage[T1]{fontenc}
\usepackage{cite}
\usepackage[pdftex]{graphicx}
\usepackage[cmex10]{amsmath}
\usepackage{color}
\usepackage{hyperref}	
\hypersetup{
	pdftitle={Simulation-based optimization of transportation costs in high pressure gas grid},
	pdfauthor={M. Kamola and S. Plamowski},
	pdfkeywords={gas transmission system} {compressor station} {gas pipeline} {layered control structure}  {simone},
}
\begin{document}
\title{Simulation-based optimization of transportation costs in high pressure gas grid}
\author{
\IEEEauthorblockN{Mariusz Kamola}\\
\IEEEauthorblockA{ Institute of Control \& Computation Engineering\\Warsaw University of Technology
\\\href{mailto:M.Kamola@ia.pw.edu.pl}{M.Kamola@ia.pw.edu.pl}}\\\vspace*{1em}
\and
\IEEEauthorblockN{Sebastian Plamowski}\\
\IEEEauthorblockA{Emerson Process Management Power \& Water Solution
\\\href{mailto:Sebastian.Plamowski@Emerson.com}{Sebastian.Plamowski@Emerson.com}}
}
\maketitle
\begin{abstract}
Design, architecture and deployment details of a~decision support system engineered to minimize operating costs of compressor stations in a gas network are presented. The system employs standard simulation software for pipelines, combined with well known optimization routine for finding optimal station control profiles in a repetitive way. A list of custom improvements is presented that make the system capable and robust enough to perform the optimization tasks. Implementation process is described in detail, covering the case of handling extra optimality criteria postulated by the user. Benefits from using the system and lessons learned are presented in the conclusions section.
\end{abstract}
\section{Introduction}
\label{sec:introduction}
When looking at the scale of investments in gas pipelines, especially those connecting Asia and Europe, it becomes obvious that this kind of fuel will remain for long a vital power medium. Evidently, growing demand for gas results from growing electric energy needs, where gas turbines can supply power in peak hours quickly. Thus, new gas sources (Nord Stream, shale gas) are natural allies of smart electric grid ideas. Gas is apparently better medium to manage than energy, at least as far as storage is concerned. However, maintenance of gas transit pipelines is as demanding as maintenance of electric trunk tracts. Equally, the daily control of the gas system is nontrivial: one must keep gas pressure, temperature and humidity within hard limits --- assuring gas quality analogously to electric current quality.

The main goal when running a gas network is to minimize its operational expenses, which are mainly the costs of pumping the gas through compressor stations. To automate the process of optimal compressor stations scheduling, one needs to possess several elements --- and to make them work together. They are:
\begin{itemize}
\item supervisory control and data acquisition (SCADA) system, to collect current network state information;
\item forecast system, to foresee network uncontrollable inputs and disturbances;
\item modeling system, to assess system behavior under assumed control;
\item optimization algorithm, to automate the process of control continuous improvement.
\end{itemize}
Here we present a complete case of such simulation optimization framework design and deployment in a region with population near 3,000,000. Further in this section we formulate the optimization problem and make an overview of ongoing similar research. In Sec.~\ref{sec:model} we present the numerical model of gas network and outline major problems while applying it in our case. In Sec.~\ref{sec:optimization} the choice of proper optimization routine is discussed. Implementation details as well as the encountered preconditions and difficulties are presented in Sec.~\ref{sec:implementation}. We conclude in Sec.~\ref{sec:conclusion} with remarks from both engineering and managerial viewpoints.

\subsection{The network}\label{sec:network}
\begin{figure*}[tb]
\begin{center}
	\includegraphics[trim=0mm 0mm 55mm 0mm,clip,height=\textwidth,angle=270]{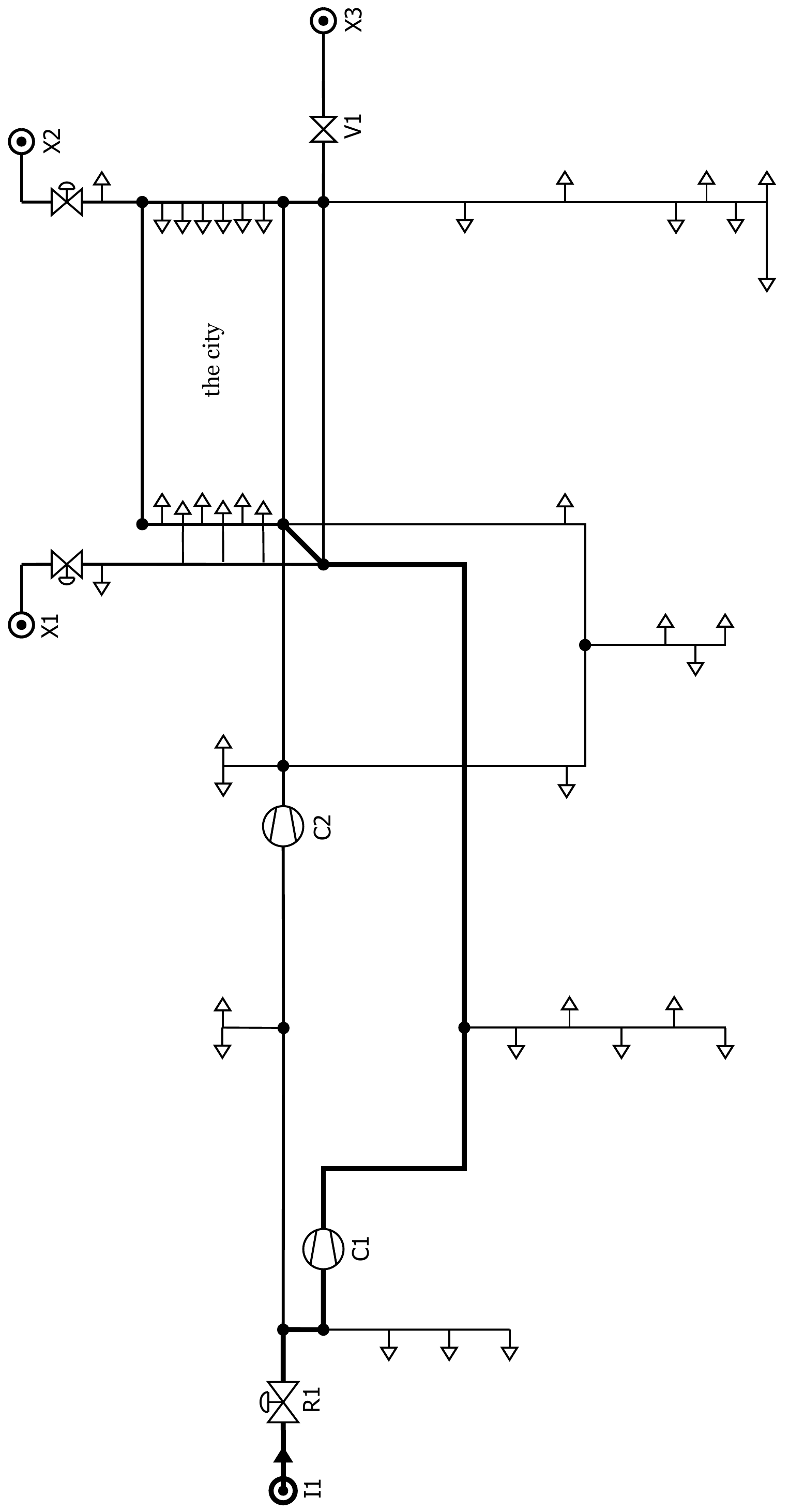}
\caption{Simplified connection topology of the gas network. Line widths reflect pipe diameters and lengths. Sideway triangles stand for retail outlets. Symbols meaning: {\tt I1} --- the major intake, {\tt R1} ---  the major reduction valve, {\tt C1,2} --- compressor stations, {\tt X1,2,3} --- interconnection points with their automatic or simple ({\tt V1}) valves.}\label{fig:gswro-all}
\end{center}
\end{figure*}
Structure of the pipeline network being subject to control is presented in Fig.~\ref{fig:gswro-all}. This is a simplified graph of regional gas network in one of Polish regions, governed by the regional gas dispatching division. The system operates at pressure up to 8 MPa, with its two main roles: to guarantee reception of gas volume prescribed by the national dispatcher from intake {\tt I1}, and to supply the region's capital (population 700,000). The system is connected with the rest of the national grid by three interconnection points, supplied either with controlled or ordinary valves. Gas surplus from {\tt I1} is forwarded to {\tt X1}, {\tt X2} and {\tt X1} in summer, to be accumulated in underground storage. In winter, gas deficit in the region is balanced by importing it the reverse way. In autumn and spring flow alternation at interconnection points may follow daily patterns.

Depending on situation (especially on regional demand and gas provider's pressure at {\tt I1}), two, one or none of compressor stations may be operational. Gas bypasses inactive stations (this piece of piping is not shown in the figure), pushed by pressure differences. Despite lack of dedicated gas storage, some amount of gas can be accumulated just in pipelines (consider that the typical pressure in network is 4 MPa, vs. the maximum of 8 MPa) thus helping in short-term balancing local demand and flows at {\tt I1}, {\tt X1}, {\tt X2} and {\tt X3} --- but this is rather expensive way to operate the network.

It should be noted that the thickest pipeline, served by {\tt C1} is currently the main route in the system, although it takes substantially longer for gas to reach the main city and the exchange points located even further. This is because of the mere diameter of the pipe, being nearly twice as thick as {\tt C2} route. Consequently, in favorable circumstances {\tt C2} is to be switched off first, and sometimes {\tt C1} can be put in bypass mode as well.

\subsection{Problem formulation}\label{sec:problem}
The main objective for a~gas network operator is to keep the total fuel consumption by compressor stations $Q$ to its possible minimum allowed by system constraints. Let us consider this consumption in time horizon $\Delta T$ as an integral
\begin{equation}\label{eq:goal}
	Q(t_0)=\int_{t=t_0}^{t_0+\Delta T}\sum_{i=1}^M Q_i(f_i(t),h_i(t),p_i(t))dt
\end{equation}
where $Q_i(f_i(t),h_i(t),p_i(t))$ denotes fuel consumption at time~$t$ by compressor station~$i$ as a~function of gas flow $f_i(t)$ through the station, of the compression ratio $h_i(t)$ and of input pressure $p_i(t)$.  For each compressor station $i\in\{1,...,M\}$, gas flow is the control variable. Since the control is discretized evenly at $N$ time moments $t_0,\;t_1=t_0+\delta t,...,\;t_{N-1}=t_0+(N-1)\delta t$, we can define a vector of $MN$ decision variables as
\begin{equation}\label{eq:control}
\begin{split}
	{\bf f}= [ & f_1(t_0),...,f_M(t_0),..., \\
		     & f_1(t_{N-1}),...,f_M(t_{N-1})]\;\;\;.
\end{split}
\end{equation}
To evaluate (\ref{eq:goal}), one needs to know the trajectories ${\bf c}(t)$ and ${\bf p}(t)$. They must be computed by gas network simulation software. Both $\bf c$ and $\bf p$ belong to a set of dependent variables, $V=\{{\bf c}, {\bf p}, {\bf y}\}$, where $\bf y$ stands for a~vector of any system state variable other than pressure ratio and input pressure. Trajectories of elements in~$V$ describe fully response of the system to the control. However, to calculate the response, the initial system state $V(t_0)$ and any other uncontrolled input trajectory must be provided. If we denote the uncontrolled inputs by vector~$\bf w$, we can write the simulation task as a~functional
\begin{equation}\label{eq:simulation}
	V(t)=S(V(t_0),{\bf f},{\bf w}(t))\;\;\;.
\end{equation}
There are three main kinds of uncontrolled inputs: gas intakes or outflows, pressure values in transit pipes, and gas or air temperatures. The way they are forecast varies; it is described in detail in Sec.~\ref{sec:model}. Here we can assume all values except~$\bf f$ as time continuous, as their granularity is an order of magnitude smaller than the control discretization period~$\delta t$.

The optimization problem is to find an optimal control sequence~$\bf f^\star$ defined as in (\ref{eq:control}) minimizing the goal function as defined in (\ref{eq:goal}), with the simulation process (\ref{eq:simulation})  considered as equality constraint. Additionally, the problem must be solved subject to box constraints on control as well as on every dependent variable:
\begin{eqnarray}
	\label{eq:xbox}f_{i,\min}\le f_i\le f_{i,\max}\;,&\\
	\label{eq:ybox}\forall_{t\in <t_0,t_0+\Delta t>}\;:&   \;{\bf c}\in<{\bf c}_{\min},{\bf c}_{\max}>\;,\\
					& {\bf p}\in<{\bf p}_{\min},{\bf p}_{\max}>\;,\nonumber\\
					& {\bf y}\in<{\bf y}_{\min},{\bf y}_{\max}>\;.\nonumber
\end{eqnarray}
Please note the different roles of the constraints. For example, minimum pressures in remote parts of network graph provide guarantees of gas provisioning capacity for retail customers. The minimum values for pressure drops on controlled valves and compressor stations assure their proper operation and, indirectly, make it possible to use flows rather than more primitive concepts as control variables. Finally, keeping gas temperature within limits means providing system security, thus preventing overheating or even ignition.

\subsection{Similar work}\label{sec:similar}
Since the practical case presented in this paper combines usage of particular set of techniques for forecasting, numerical modeling and optimization, whilst referring to similar works by others we will try to present research results most similar in all those aspects. For broader list of alternative techniques, please refer to relevant sections on modeling and optimization.

The problem of finding optimal compressor station load profiles has been addressed already in 80s \cite{Marques88} by harnessing a flow simulator of some sort combined with sequential quadratic programming (SQP). Such approach fits the simulation-optimization scheme in general, however it relies on assumptions of smoothness of modeling functions, in order to make SQP operate effectively. Later, by introduction of optimization methods based on duality, constraints on dependent system variables could be treated effectively \cite{Osiadacz94}. While this approach was being further developed to support large-scale systems by means of hierarchical coordination \cite{Osiadacz98}, others (cf. eg. \cite{Mahlke07} and references to prior works therein) focused on other difficulties of the optimal control problem: discontinuity of modeling functions and presence of discrete decision variables. To handle that, they resorted to using non-gradient global methods (like simulated annealing) or to problem reformulation into a continuous one.

There is still much research activity going on as the pipeline infrastructure undergoes continuous intensive extensions. When the system is treated as a whole, as eg. Chinese grid presented by \cite{Jin10}, one realizes that the problem tends to be a combinatorial one. This is because a compressor is most effective in a narrow range of its working states, and finding optimal routing for gas transportation, i.e. such that compressors are either off or working at fixed point becomes a demanding task (resembling energy-optimal routing in wired computer networks --- cf. eg. \cite{ens12}). A comprehensive and up-to-date survey of problem solutions proposed in the literature has been provided by  \cite{RiosMercado12}.

\section{Modeling the gas network}
\label{sec:model}
Gas network simulator \href{http://www.simone.eu}{\em Simone} has been used for calculating static and dynamic system response to control in the case presented in this paper. There exist a number of similar packages: \href{http://www.atmosi.com}{\em Atmos Sim}, \href{http://www.b3pe.com}{\em GasWorks} suite, \href{http://www.gl-nobledenton.com}{\em Stoner Pipeline Simulator}, \href{http://www.flowmaster.com}{\em Flowmaster} and \href{http://www.stctrl.com}{\em OptiRamp}, to point out a few. They all share a~set of common functionalities: static and dynamic gas network simulation, and a set of accompanying simulation-based optimization tools supporting network design as well as optimal control. Simone simulator has been adopted in the presented case because it has been already used widely and frequently by Polish pipeline operators in engineering processes. This is an important reason since network structures evolve continuously, and maintaining two separate kinds of system models would be costly and error-prone. Having experts from the field at hand at the stage of our software development, testing and deployment was invaluable as well.

Simone basic pipeline simulation models are presented clearly by \cite{SimoneBasic04}: partial differential equations are applied to force continuity and balance momentum of flows, and Hofer formula is used by default to model friction. Gas state is determined by two-parameter second order Papay formula by default, but the choice of other approaches is up to the user.  Other rules determine gas quality, gas mixing process etc. Gas heat dynamics is modeled using fixed exponential model. The basic model of reciprocating compressor includes linear formulae for volumetric flow rate and for the torque momentum. It is also possible to perform advanced configuration of compressor stations \cite{SimoneCS07}. 
\begin{figure*}[tb]
\begin{center}
	\includegraphics[height=0.55\textwidth,angle=270]{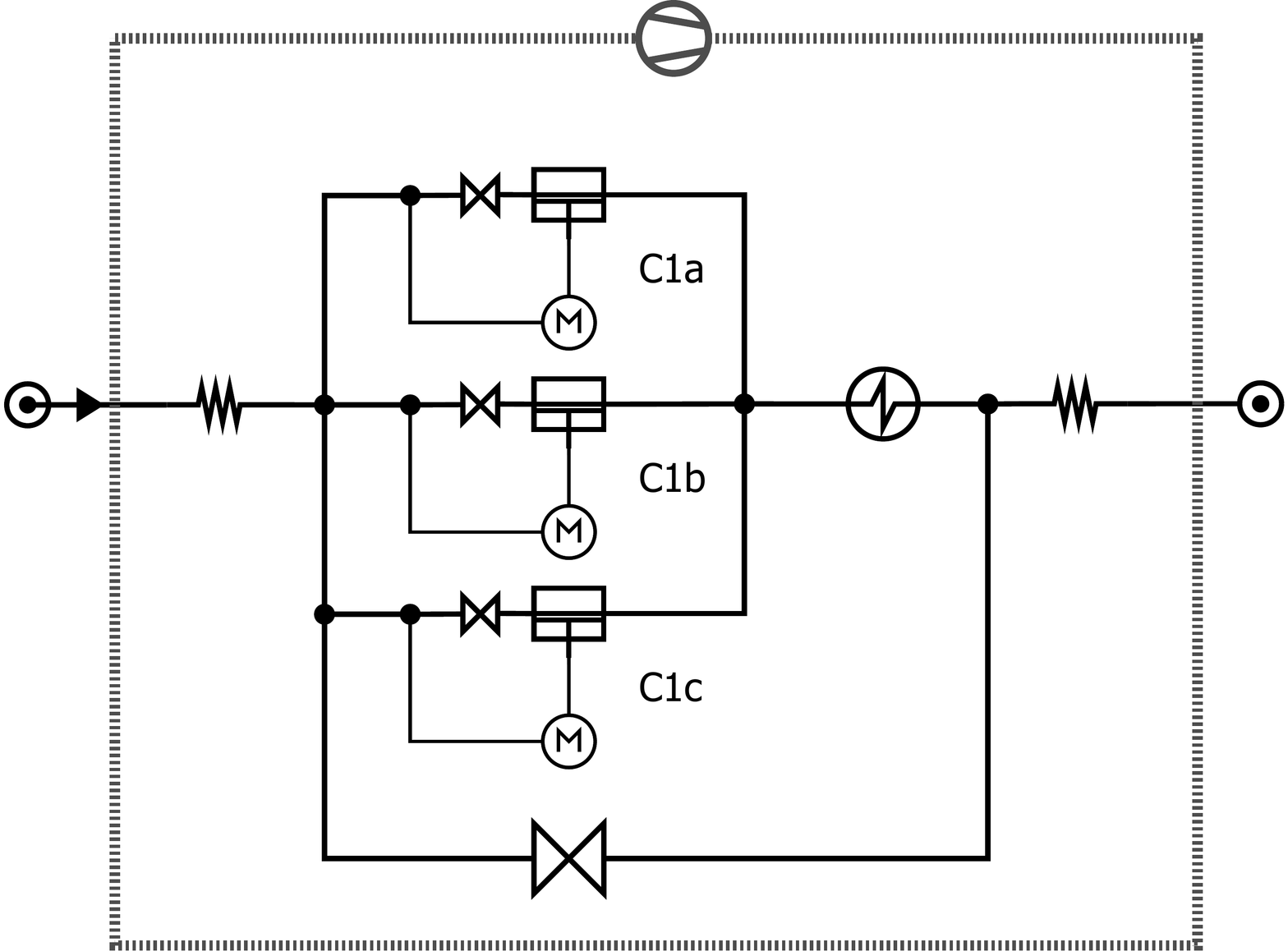}
\caption{Actual internal model of a compressor station. Three reciprocating compressors are powered by gas engines, and can be selectively switched off. Also, intake and discharge resistances, the gas cooler and the bypass valve are shown.}\label{fig:gswro-cs}
\end{center}
\end{figure*}
The detailed model of a compressor station shown in Fig.~\ref{fig:gswro-cs} presents the way one stage of gas compression is modeled in Simone. It matches exactly the actual structure of both compressing stations in the real object of interest. One of the simplest advanced station configuration options include specification of bounds for input and output pressures, and also for pressure differences that precondition station operation. Shall those be violated, the whole station is set into either off or bypass mode. More advanced options, like torque revolutions range, maximum compression ratio or empirical efficiency per every reciprocating compressor are available only under extended Simone license, that was not planned to be purchased for this project. This has become one of major reasons for which pumping costs calculated by Simone were not taken into account while calculating actual system performance during optimization.

\subsection{Network initial state estimation}
\label{sec:estimation}
Before a projected control scenario can be fed into Simone to perform dynamic network simulation, the actual network state must be estimated. This is done in Simone by running static simulation, i.e. evaluation of network conditions in a~steady state. Once having the estimate of the steady state, one can use it as a starting point for dynamic simulation. Such approach requires minimum information from SCADA systems: if, for static simulation, we replace all active network elements (compressors or controlled valves) with intake\&outlet pairs, we obtain a modified network graph. For each disjoint part of the graph we need to provide exactly one pressure measurement, and a complete list of flows on its boundaries. Based on that, all other steady-state network variables are calculated.

Certainly, such approach has its drawbacks. Firstly, it inherently ignores transient effects that usually  take place in the network, like line packing. Consequently, running any dynamic simulation from the estimated state results in sudden initial flow peaks between network segments. Therefore, we decided to ignore the results for first time step in any dynamic simulation, for the sake of production control stability. Picking the right pressure measurement point for each subgraph for static simulation is another process that cannot be automated easily. Instead of facing a dilemma, it probably would be better to provide Simone with a set of any available (and, due to transient effects, contradictory in static sense) pressure measurements, for some more intelligent network estimation routine.

As in every practical application, a number of validations and workarounds have to be implemented in order to get simulation-optimization interface work robustly. An interesting and common sources of this interaction failures worthy mentioning here were:
\begin{itemize}
\item Switching off whole branches of network by operator, for maintenance purposes. From strategic point of view totally unimportant, this caused simulator to fail because usually no real pressure measurement was available for the segment being cut off. The solution was to provide an artificial measurement value (2.2 MPa) to Simone. This reparation procedure is applied automatically.
\item Hitting predefined network constraints in dynamic simulation caused simulation to fail without clear info about the failure reason. Parsing of simulation log is done in such case, in order to find the network element where constraints are violated, and to calculate penalty function appropriately.
\item Measuring devices maintenance resulted from time to time in their improper reconfiguration, e.g. providing measurement in different units (for flows: ${\rm Nm}^3/{\rm h}$ instead of ${\rm Nm}^3/{\rm h}\cdot1000$; for pressures $p$[MPa] instead of $(p-0.1)$[MPa]). A database with mapping between SCADA measurement points and Simone object properties was developed, and adequate bounds for measurement validation have been specified there.
\end{itemize}

\subsection{Modeling uncontrolled inputs}
\label{sec:uncontrolled}
We can roughly classify uncontrolled inputs to simulation procedure (\ref{eq:simulation}) according to their level of uncertainty. Planned flow profiles are called nominations: they are provided to the network operator by the national dispatcher unit and describe flows at major exchange points. They can be also provided (and, in fact, it tends to be so) by retail customers. Retail customers are not individuals: they are either larger businesses or retail resellers. Upon providing their accurate estimates they obtain better prices from the operator, so they have incentive to provide true estimations.

The operator must also take care about all inputs not modeled by its supervisor or customers. They are temperatures, and retail flows not used as nominations. To forecast them all, a third party modeling software is used that takes into account historical values, the calendar (with exceptions) and --- particularly for the external temperature --- weather forecasts. Currently the underlying modeling technology applied is neural networks, but other models, like autoregressive with moving average and external input (ARMAX) are also applicable. Obviously, forecasts made this way do have bigger estimation error, but still they are provided in crisp form (i.e. the forecasts do not address probability in any way).

Finally, there is one variable considered completely unpredictable yet having huge impact on the system operation: the pressure at major intake, {\tt I1}. Should it be predicted accurately, the operator could profit enormously, e.g. by packing its lines only when it is low, if the pressure rise is to happen soon. However, the major gas supplier is not obliged to provide such information, and the best practice in such case is to assume the pressure to stay at its currently observed value for the whole optimization horizon. Such strategy is, after all, no so bad: sudden pressure drops are rare; it shows rather a mild trend (e.g. 5\% rise per hour) that gets easily consumed by the applied repetitive optimization scheme.

\section{Optimization algorithm}
\label{sec:optimization}
All of the simulation packages mentioned in the begin of Sec.~\ref{sec:model}  provide built-in optimization capabilities. When tightly embedded in modeling software, an optimization procedure may profit best from information not available externally, like function derivatives. Consequently, procedures applied by software manufacturers are usually gradient ones, like SQP or SLP. However, if discrete decision variables are to be supported, one needs to resort to global methods or problem reformulation (e.g. initial relaxation).

Eventually, the built-in optimization routine shipped with Simone was not used in the reported case for a combination of reasons:
\begin{itemize}
\item Optimization support in Simone comes as an extra paid option, for which neither network operator nor the automated decision support system implementor was prepared.
\item On-line data acquisition from the system through a set of Simone add-ons is required so that the model is fed with actual measurements of the system. However, no driver development was planned, and it was found more convenient to load actual system state information into Simone via startup parameters as scenarios and initial conditions.
\item According to project scope definition, the software should {\em assist} network operator with decisions taken in {\em current} network configuration (i.e. the set of active compressors). Therefore, the resulting optimization problem is a continuous one. The strategy for individual compressor activation is based on different rules, and was left up to the operator.
\item Preliminary tests showed that application of an external, relatively simple optimization procedure gives acceptable results, allowing at the same time better inspection of eventual reasons of optimization failure, and better error reporting to the operators.
\end{itemize}

Repetitive control algorithm was chosen for application, where Powell conjugate direction method was applied to find the optimal control each time. Powell method \cite{Powell64} belongs to a group of non-gradient optimization algorithms where line searches are initially done along unit vector directions. Later, those search base directions get replaced by a combination of steps already made, and periodic base reset is applied to avoid linear dependence of the search directions. 

The long time of objective function evaluation through simulation is  the  main problem using Powell optimization algorithm in the practical applications. The algorithm needs to estimate the value of the objective function in many points in each step – it means that simulation in Simone has to be called many times during one algorithm step. If one simulation takes about 5  sec. only 180 simulations can be carried out in 15-minute repetitive control interval. This number is definitely insufficient to find the minimum. Two mechanisms were implemented to overcome this limitation:
\begin{itemize}
\item A dedicated simulation result cache was implemented, where the objective function values are stored. In the situation where the next step of the algorithm requires value of the objective function for a nearby point to the point that has already been analyzed, the value of the objective function is taken from the buffer. 
\item If the accuracy of the found solution in one algorithm run (15 minutes at most) is insufficient, the best solution found so far is used as a starting point for the next run. Thus, even in case of the initial starting point located far from the optimum, the algorithm slowly recovers. 
\end{itemize}
These modifications reduce time of calculation to the acceptable level. Box constraints on dependent variables are handled through penalty functions (applying first or second order functions did not influence results significantly).

 Interestingly, contrary to one of authors previous experience \cite{Kamola07} simple constraints on simulated variables do not distort the search domain such that deterministic and local optimization methods cannot be used. Instead, Powell was observed to converge somewhat slowly but steadily to a unique, reasonable solution in our case.

\section{Implementation: the regulation layer}
\label{sec:implementation}
The well known specifics of the implementation and integration phases consists in interconnection of multiple technologies, some of which may not be eager to interface the others. The less known one is that the customer (gas network operator in this case) going to apply fully automated control realizes that his perception of optimality is deeply human, and that quality of his models of the system is far from sufficient.

\subsection{Multiple optimality criteria}
\label{sec:multicriteria}
We started with clear optimality criterion formulation in (\ref{eq:goal}). However, skilled human network operators express their expert knowledge and practice posing additional desirable optimization goals for the system. They sound reasonable, as in our case:
\begin{enumerate}
\item the system should keep pressure difference at valve {\tt R1} to its possible minimum;
\item the system should distribute load across compressors so that their mileage is balanced.
\end{enumerate}
Looking closely, we see that those postulates can be at least inconsistent with (\ref{eq:goal}). Moreover, postulate 2 contradicts it because fuel consumption minimization means that most effective compressors should be fully loaded all the time. Problems of similar nature have been studied recently in \cite{Aicha13}: the authors consider there a multicritreria model predictive control, comparing two methods of picking up a~unique solution out of the Pareto front.

To handle postulate 2, we decided to pass the decision which compressor should be activated into the hands of a human operator. Shall there be adequate disproportion in mileage, he is presented a suggestion to alternate between machines, and so {\em he} is responsible for weighing fuel consumption against mileage.  Postulate 1 has deeper implications. We decided to approach it by scalarizing the overall control objective. In practice this means introduction of a customized PID regulator to control pressure difference at {\tt R1} by means of calculating output pressure set points for every individual compressor. The regulator also accepts optimization results as corrections to its original control signals.

\subsection{Layered control structure}
\label{sec:layered}
\begin{figure*}[tb]
\begin{center}
	\includegraphics[height=0.75\textwidth,angle=270]{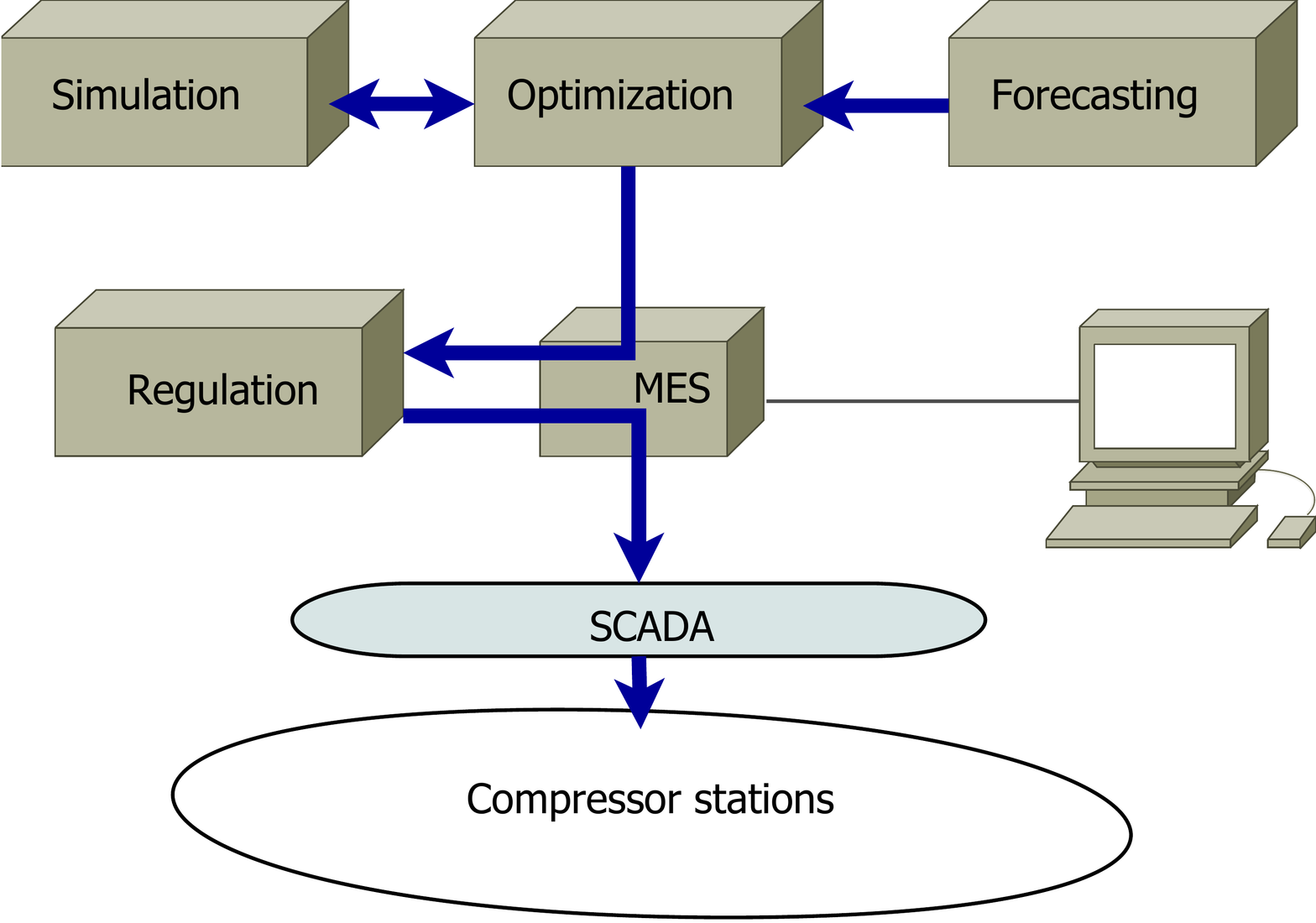}
\caption{Control blocks and control flow for the presented case. }\label{fig:hierarchy}
\end{center}
\end{figure*}
The software structure diagram presented in Fig.~\ref{fig:hierarchy} shows full flow of information between the various modules. The main element of the solution is Manufacturing Execution System (MES), which was used for data integration process and acts as a layer of storage and data transfer. Using MES as an independent platform allowed to separate operational SCADA data from data optimization computational tasks so that operation of system optimization of has no effect on the efficiency of the SCADA system.

MES system works with other subsystem to take data for optimization calculations. Low frequency data (like flow and pressure) are read from upper SCADA system. Data rapidly changing, such as torque rotation, the pressure before and after the compressor and the temperature of radiators are collected via Modbus server directly from compressor station SCADA. All data are collected in MES system and they are anytime available for computing modules.

Data about flows are used by the forecasting module, which determines daily profiles based on the value history. The profiles are scaled by the current and forecast values of the temperature. In addition, the learning mechanism has been used in the  algorithm, so that the profile can adapt to changing conditions such as the acquisition of new customers or changes in  customer behavior due to economic factors. The profiles are stored in MES database  where they are available for simulator and optimization modules.

Optimization module performs the task of optimization (described in Sec.~\ref{sec:optimization}), which boils down to the calculation of the optimal trajectory controls (vectors of the compressor flows)  basing on the simulator network model, forecast gas consumption, the current situation in the network and operating point of compressors. Optimization calculations are carried out in a way that guarantees the following technical requirements:
\begin{itemize}
\item working in on-line mode with on-line data, taking into account the possibility of uncertain data;
\item calculations are carried out on a full-featured network scheme currently used by gas grid operator to perform simulations on-line;
\item application presents an optimal (best found) variant with regard to any restrictions imposed by the operator and technical parameters of the transmission system, the results are shown on each point in a graphical form, the dispatcher after receiving the results has the ability to track changes in pressure and flow at each measurement point in advance, for at least 24 hours ahead.
\end{itemize}

\subsection{Regulation layer}
\label{sec:regulation}
The results of calculations made by the upper control layer are transferred to the lower control layer. The lower control layer in the described solution is the direct control layer, which task is running (once every 10 s) to correct the values calculated by the optimization module (upper control layer). It  is based on the current measurement of:
\begin{itemize}
\item compressor inlet pressure,
\item compressor output pressure,
\item machines speed,
\item gas temperature for cooler,
\item pressure at the main intake {\tt I1},
\item pressure after the controlled valve  {\tt R1}.
\end{itemize}
Another task of the lower control level is to maintain equal load of machines in compressor stations. Input data are taken directly from the MES system, the calculated values are transferred also to MES system, where through the  Modbus server they are sent to the compressor station SCADA as valid controls.

\section{Conclusion}
\label{sec:conclusion}
A real case of simulation-optimization approach to accomplish fully automated network controller has been presented here. The system was deployed in 2009 and redesigned after major topology changes in 2012. Application of the system resulted in about 10\% reduce of gas consumption used by compressors.

We do not claim that the optimization procedure or the modeling of compressors, or even the forecasting methods applied are in the forefront of research in the field. However, applying robust and understandable  approaches is bigger merit in practical cases than reaching for sophisticated solutions. End user prefers application robustness and exhaustive diagnostics capabilities over a routine which is capable of finding global optimum at high computation cost.

It is important to remember that user requirements are often formulated spontaneously, and express all activities done so far manually by the experts. The requirements, when stated formally, may turn out to be contradictory: handling such cases by scalarization or by involving human decision are possible solutions.

Finally, it must be said that profits from decision support deployment are not limited to monetary savings, tidying up system inventory or making think about control goals in a systematic and rational way. A good automated control system is also a win-win solution for parties with formerly conflicting interests --- as the management and dispatching teams in our example. The former cared for fuel savings rigorously as it determined the company valuation; the latter kept the system far from constraints in order to minimize risk of failing to meet transit agreements, and to have some spare linepack in case of a failure. With the control software properly adjusted both can enjoy their needs being satisfied.

\section*{Acknowledgments}
\label{sec:ack}
The work reported here was carried out within a commercial contract with the mentioned gas network operator, where both authors were employed as external consultants.

\bibliographystyle{IEEEtran}
\bibliography{IEEEabrv,gswro}

\end{document}